\documentclass{article}
\usepackage{epsfig}
\usepackage{aaspp4}

\newcommand{\Lsun}{\mbox{$L_{\odot}$}}
\newcommand{\Msun}{\mbox{$M_{\odot}$}}

\newcommand{\Lstar}{\mbox{$L_*$}}
\newcommand{\Mstar}{\mbox{$M_*$}}
\newcommand{\Rstar}{\mbox{$R_*$}}

\newcommand{\etal}{\mbox{et~al.}}
\newcommand{\Teff}{\mbox{$T_{\rm eff}$}}

\newcommand{\Msunyr}{\mbox{$M_{\odot}~yr^{-1}$}}
\newcommand{\Msunyear}{\mbox{$M_{\odot}~yr^{-1}$}}

\newcommand{\eg}{\mbox{\it e.g., }}

\begin{document}


\title{Ongoing massive star formation in the bulge of M51 
      \footnote{
    Based on observations with the NASA/ESA Hubble Space Telescope,
    obtained at the Space Telescope Science Institute, which is
    operated by AURA, Inc., under NASA contract NAS 5-26555}
      }

\author{ H.J.G.L.M. Lamers  \altaffilmark{1,2}
         N. Panagia     \altaffilmark{3,4}
         S. Scuderi     \altaffilmark{5}
         M. Romaniello  \altaffilmark{6}
         M. Spaans      \altaffilmark{7}
         W.J. de Wit    \altaffilmark{1,2}
         R. Kirshner    \altaffilmark{8}}


\altaffiltext{1} {Astronomical Institute, Princetonplein 5, NL-3584 CC Utrecht,
                 The Netherlands; {\tt lamers@astro.uu.nl} and 
                 {\tt w.j.m.dewit@astro.uu.nl}}
\altaffiltext{2} {SRON Laboratory for Space Research, Sorbonnelaan 2, 
                 NL-3584 CA Utrecht, The Netherlands}
\altaffiltext{3} {Space Telescope Science Institute, 3700 San Martin Drive,
                 Baltimore, MD21218, USA;
                 {\tt panagia@stsci.edu} }
\altaffiltext{4} {On assignment from the Space Science Department of ESA}
\altaffiltext{5} {Osservatorio Astrofisico di Catania, Italy;
                 {\tt sscuderi@alpha4.ct.astro.it} }
\altaffiltext{6} {European Southern Observatory, Karl-Schwarzschild Strasse 2,
                  Garching-bei-Muenchen, D-85748, Germany;
                 {\tt mromanie@eso.org}}
\altaffiltext{7} {Kapteyn Astronomical Institute, University of
                 Groningen, PO Box 800, NL-9700 AV Groningen, The Netherlands; 
                 {\tt spaans@astro.rug.nl}}
\altaffiltext{8} {Harvard-Smithonian Center for Astrophysics, 60 Garden Street,
                  MS-19, Cambridge, MA 02138, USA; 
                 {\tt rkirshner@cfa.harvard.edu}}

\date{Received date, accepted date}

\markboth{Massive star formation in the bulge of M51}{}

\today


\begin{abstract}
We present a study of $HST-WFPC2$ observations of the inner kpc of the
interacting galaxy M51 in six bands from 2550 \AA\ to 8140 \AA. The 
images show  an oval shaped area (which we call ``bulge") of about
$11 \times 16$ arcsec or $450 \times 650$ pc around the nucleus  that
is dominated by a smooth ``yellow/reddish" background population with
overimposed dust lanes. These dust lanes are the inner extensions of
the spiral arms.   The extinction properties, derived  in four fields
in and outside dust lanes, is similar to the Galactic extinction law.
The reddish stellar population has an intrinsic color of $(B-V)_0
\simeq 1.0 $ suggesting an age in excess of 5 Gyrs.

We found 30 bright point-like sources in the bulge of of M51 i.e.
within 110 to 350 pc from the nucleus. The point sources have $21.4 < V
< 24.3$, many of which are  blue with  $ B-V < 0$ and are bright in the
UV with $19.8 < m_{2550}< 22.0$. These objects appear to be located in
elongated ``strings" which follow the general pattern of the dust lanes
around the nucleus.  The spectral energy distributions of the
point-like sources are compared with those predicted for models of
clusters or single stars. There are three reasons to conclude that
most of  these point sources are isolated massive stars (or very
small  groups of a few isolated massive stars) rather than clusters:\\
(a) The energy distributions of most objects are best fitted with
models of  single stars of $M_V$ between -6.1 and -9.1, temperatures 
between 4000 and 50000 K, and with 4.2 $<$ log $L/\Lsun <$ 7.2, and  $12
< \Mstar < 200~ \Msun$. \\
(b) In the HR diagram the sources follow the Humphreys-Davidson 
luminosity upper limit for massive stars. \\
(c) The distribution of the sources in the HR diagram shows a gap in
the range of $20~000 < \Teff < 10~000$ K, which agrees with the rapid
crossing of the HRD by stars, but not of clusters.\\ We have derived
upper limits to the total mass of lower mass stars ($\Mstar < 10~
\Msun$), that could be ``hiding'' within the point sources. For the
``bluest'' sources the upper limit is only a few hundred \Msun.

We conclude that the formation of massive stars outside clusters
(or in very low mass clusters) is occurring in the bulge of M51.

The estimated star formation rate in the bulge of M51  is 1 to $2
\times 10^{-3}$ \Msunyr, depending on the adopted IMF. With the
observed total amount of gas in the bulge, $\sim 4 \times 10^5$ \Msun,
and the observed normal gas to dust ratio of $\sim 150$, this  star
formation rate could be sustained for about 2 to $4 \times 10^8$ years.
This suggests that the ongoing massive star formation in the bulge of
M51 is fed/triggered by the interaction with its companion about $4
\times 10^8$ years ago. 
The star formation in the bulge of M51 ia compared with tat in bulges
of other spirals.

Theoretical predictions of star formation suggest that isolated massive
stars might be formed in clouds in which 
H$_2$, [OI] 63 $\mu$m and [CII] 158 $\mu$m are the dominant coolants.
This is expected to occur in regions of rather low optical depth, $A_V
\le  1$, with a hot source that can dissociate the CO molecules. These
conditions are met in the bulge of M51, where the extinction is low
and where CO can be destroyed by the radiation from the bright 
nuclear starburst cluster in the center. 
The mode of formation of massive stars in the bulge of M51  
may resemble the star formation in the early Universe, when the
CO and dust contents were low due to the low metallicity.

\keywords{ 
         galaxies: bulge --
         galaxies: spiral --
         galaxies: interaction --
         stars: formation --
         stars: evolution --
         }

\end{abstract}

\section{Introduction}

The spiral galaxy M51 (NGC 5194, the Whirlpool nebula)  and its
peculiar companion galaxy NGC 5195 form a typical example of galaxy
interactions.  After the pioneering hydrodynamical simulations by
Toomre and Toomre (1972), several authors have tried to explain the
grand design spiral shape and the tidal arms of this interacting
system. Hernquist (1990) and Barnes  (1998) have critically
discussed the successes and the problems of explaining the morphology
of the NGC 5194/5195 system, in particular  the radial velocities of
the two galaxies, the two-armed spiral structure of M51, the connecting 
tidal arm and  the large H\,{\sc i} arm. The best model is found for a
relative orbit that is almost in the plane of M51, for a mass ratio of
NGC 5194/5195 $\simeq 2$, a pericenter distance of  17 to 20 kpc and a
time since pericenter of $2.5\times  10^8 <t < 4\times 10^8$ yrs 
(Barnes 1998).

The NGC 5194/5195 system is ideally suited for the study of triggered
star formation due to galaxy-galaxy interactions. For this reason we
have started a series of studies on the different aspects of star
formation in M51, based on $HST-WFPC2$ observations in six broadband
filters. The nucleus was studied by Scuderi et al. (2001; hereafter
called Paper I), who found that it contains a total stellar mass of
about $2\times 10^7$ \Msun\ within the central 17 pc and a bright point
source of $2\times 10^6$ \Msun\ within the inner 2 pc.  In this paper
we study the properties of the elongated  ``yellow/reddish" region
around the nucleus, hereafter called ``the bulge'',  and of 30 bright
point-like sources that we discovered in it. These point sources
indicate ongoing star formation in the bulge region, which is otherwise
dominated by old ($>$ 5 Gyrs) stars (Paper I). 

The full $HST-WFPC2$ image of M51 is published is Paper I. The image
shows that the nucleus is surrounded by an elongated bulge of about 460
$\times$ 860 pc that is dominated by an old stellar population.   The
spiral arms containing H\,II regions start outside the bulge. We adopt
a distance of $d=8.4 \pm 0.6$ kpc, corresponding to a distance modulus
of $29.62 \pm 0.16$,  which is based on the brightness distribution of
planetary nebulae (Feldmeier, Ciardullo \& Jacoby 1997). At this
distance, 1 arcsec corresponds to a linear distance of 40.7 pc, one
$HST-PC$ pixel of 0.046$^"$ corresponds to 1.87 pc and an $HST-WFC$
pixel of 0.1$^"$ corresponds  to 4.1 pc.

In \S~2 we describe the observations and the data reduction. 
In \S~3 we discuss the interstellar extinction in the bulge.
In \S~4 we describe the properties of 30 bright point-like sources
in this region. 
In \S~5 their energy distributions are compared with those
predicted for  clusters with different ages and mass, and for single
stars of different effective temperatures and radii.  We will show that
most of them are  very luminous young stars  (single or multiple),
rather than clusters.
The star formation rate in the bulge of M51 is derived in \S~6.
In \S~7  we compare the formation of massive stars in the bulge of M51 with
clusters near the Galactic center, in the
interaction region of the Antennae galaxies, and in the bulges of
other spiral galaxies.  
We also discuss the predicted mode of star formation under the
conditions that prevail near the nucleus of M51.
 The conclusions are given in \S~8.


\section{Observations and Data Reduction}

M51 was observed with $HST-WFPC2$ on May 12, 1994 and on January 15, 
1995 as part of the $HST$ Supernova Intensive Study (SINS) program to
study SN 1994I (\eg Millard \etal\ 1999).  The galaxy was observed
through the wide band filters F255W  and F336W in 1994 and through the
wide band filters F439W,  F555W, F675W and F814W in 1995. We will refer
to these filters  as the $UV$, $U$, $B$, $V$, $R$, and $I$ filters. 
The observations in the $UV$, $U$ and $B$ filters  were split into
four, three and two exposures of 500, 400, 700 $s$  respectively, while
a single exposure of 600 $s$ was taken with the  remaining filters.

The data were processed through the PODPS  (Post Observation Data
Processing System) pipeline for bias  removal and flat fielding. The
removal of the cosmic rays in the $UV$, $U$ and $B$ images was
accomplished by combining the available exposures. For the images 
taken in the other three filters we used a simple procedure, which
consists in combining images obtained in adjacent bands, that  allowed
us to remove most of the contribution from the cosmic rays (see Paper
1). Unfortunately the corrected images still showed some residual
contamination from cosmic rays which  could have affected the
photometry. To obtain a list of the point-like sources in each filter
we used an unsharp masking method. This was done because of the
complexity of  the background emission. In fact, it seems that most of
the point-like  sources lie on or close to dust lanes that are the
inner extensions of the spiral arms structures. The method consisted in
smoothing each image using a Gaussian with 4 pixels FWHM and then
subtracting the smoothed image from the original image. From each image
the list of sources was obtained by picking up all the objects above a
threshold, variable from about 3 to 13 times the local value of the
background depending on the filter.  These lists contained also
residual cosmic rays so we ``cross--correlated" them, selecting only
those objects that were present in at least two different filters. 

This procedure was  applied only to the optical filters. The $UV$ and
$U$  images were taken 8 months earlier than the optical images, which 
means that the orientation of the spacecraft was not the same in the
two epochs. In the 1995 observations the nucleus of M51 was near the
center of the PC-images, but in  1994 the images were centered on
SN~1994I and the nucleus was at the edge of the PC-image. This means
that the $UV$ and $U$ observations of  about half of the point sources
were made with the PC-camera and the rest were made with the WF-camera.

To obtain the photometry of the point-like sources in the $UV$ and  $U$
filters we first took the list containing the positions (pixel
coordinates)  of the point-like sources in the optical images  and
rotated the image, taking the position of SN 1994I  as origin, to match
the different orientation.  In this way we identified the same
point-like sources in the $UV$  images as in the optical images.

We performed aperture photometry using an aperture radius of 2 pixels
and calculating the sky background in an annulus with internal and
external radius of 5 and 8 pixels respectively. The correction for the
aperture was obtained from a theoretical PSF obtained with Tiny Tim
(Krist \& Burrows 1994). 
 The flux calibration was obtained using the
internal calibration of  the $WFPC2$, using the spectrum of Vega as
photometric zero-point (Whitmore 1995). The uncertainty in the
photometry was computed by taking into account photon noise,
background noise and CCD read-out noise only. Charge transfer and distortion
effects were not taken into account. Tests showed that these effects have
little influence on the colours (less than about 0.03
magn) and on the magnitudes (less than 0.04 magn).

\begin{figure}
\centerline{\psfig{figure=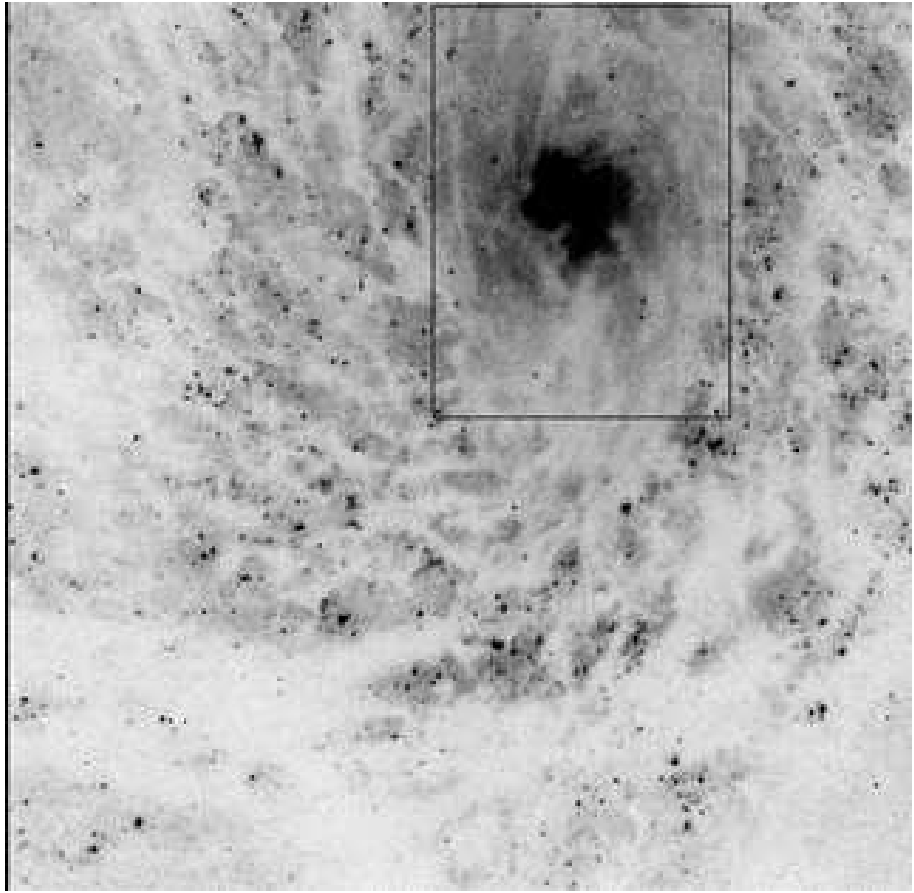,width=12cm}}
\caption[]{The negative $F439W$ image of M51 taken with the $HST$ 
Planetary Camera.
The field is $36.8 \times 36.7$ arcsec or $1.48 \times 1.48$ kpc.
The nucleus and the spiral arms are clearly visible. 
The rectangle indicates the ``bulge region" that is enlarged in Figure 2. 
The dust lanes in the bulge follow a spiral structure as if 
they are the inner extensions of the spiral arms seen at larger distance.}
\label{fig:pcimage}
\end{figure}

\begin{figure}
\centerline{\psfig{figure=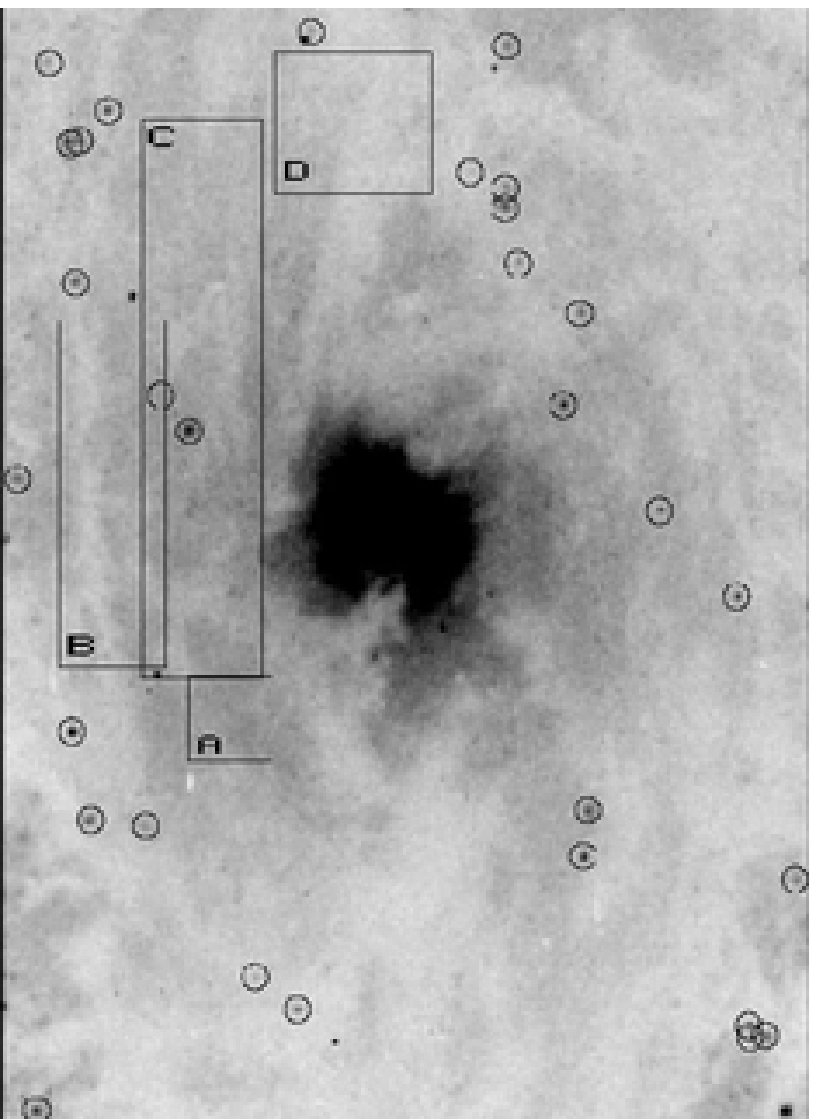,width=12cm}}
\caption[]{The negative $F439W$ image of the region of $11 \times 16$ arcsec or
$450 \times 650$ pc around the nucleus of M51 observed with the $HST-PC$-camera.
Apart from the nucleus and the dust lanes, the region has a smooth
structure with colors indicating an old, $> 5\times 10^9$ yrs, population. 
The four fields, used for studying the extinction
curve  and the location of the ``bulge point sources" are indicated.}
\label{fig:bulge}
\end{figure}

Figure \ref{fig:pcimage} shows the $B$-image taken with the
Planetary Camera. The bulge region is indicated.
Figure \ref{fig:bulge} shows the region of the bulge
enlarged. The symbols and the marked areas are described below.

%

\section{The interstellar extinction}

The region around the nucleus of M51, in an area of  about 11 $\times$
16 arcsec = 460 $\times$ 680 pc,  appears ``yellow/reddish" in the 
$HST$ images.  The light from this region, the bulge, is dominated by an
old stellar population.  The brightness distribution in the bulge is
smooth and homogeneous  in color apart from the many darker ``lanes"
which are obviously due to dust extinction, see Fig.
\ref{fig:bulge}.   Assuming a homogeneous stellar distribution in
color and magnitudes, the  extinction can be studied by comparing the
magnitudes and colors of the dusty regions with those in adjacent
regions were the extinction appears to be small. 

\begin{figure}
\label{fig:absebv}
\centerline{\psfig{figure=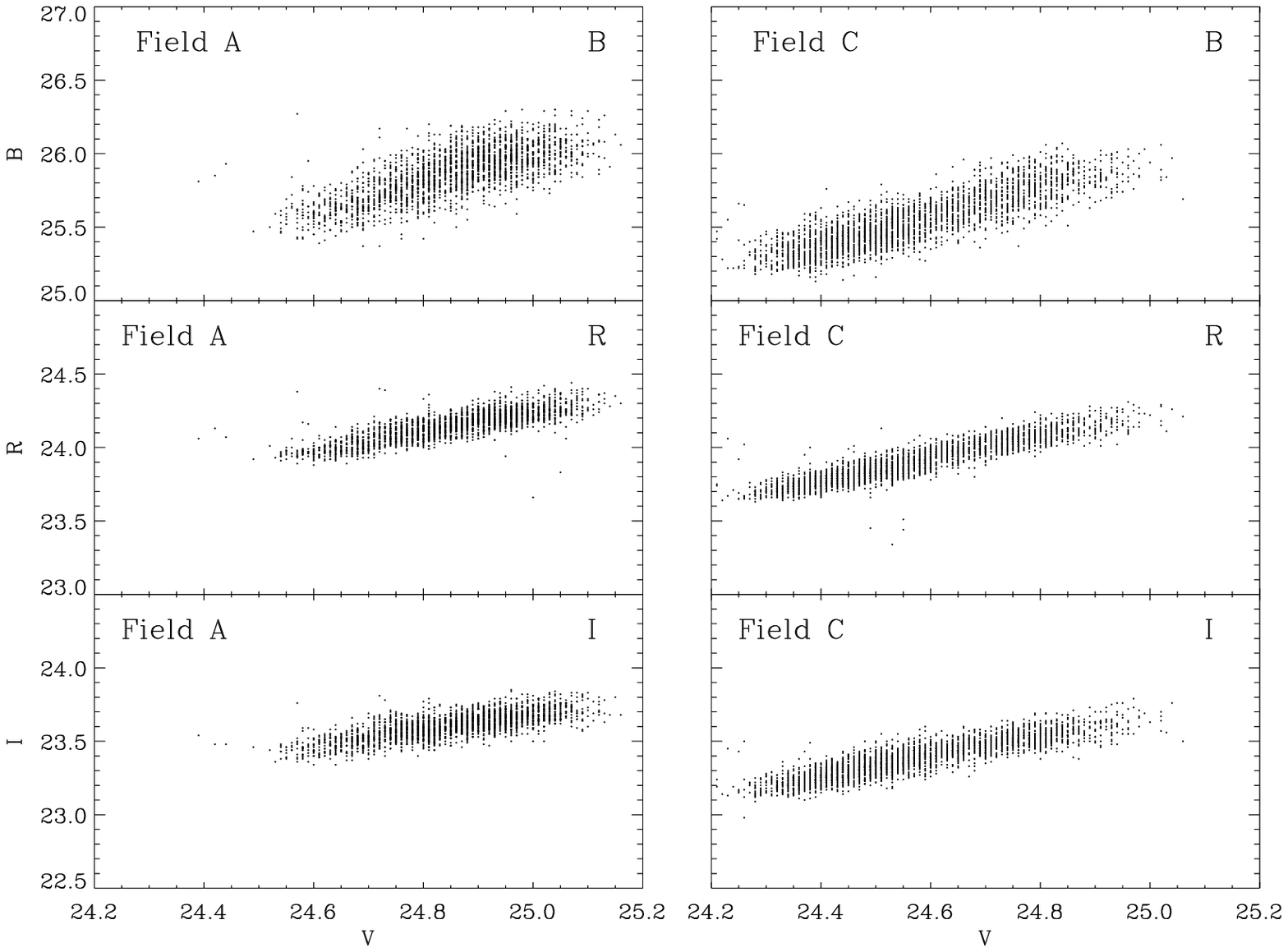,width=14.0cm}}
\caption[]{The $m_\lambda$ versus $V$ plots (in the Vega-magnitude
sytem) for the $B$, $R$ and $I$ bands of the fields A (left)
and C (right) of the bulge used for the study of the extinction 
law in the bulge. The slope of these relations define the
extinction ratios $A_{\lambda}/A_V$.}
\end{figure}

To this purpose we have selected in the bulge four areas  with
different extinctions (Fields A to D).  The locations of these
fields are indicated in Fig. \ref{fig:bulge}. 
For each of the fields we constructed  $m_{\lambda}$ versus $V$ and 
color-color plots of all the pixels. Figure \ref{fig:absebv}
shows the $m_{\lambda}$ versus $V$ plots for fields A and C.
Assuming that the background population of bulge stars has a uniform colour
(see Paper I), the slopes of the relations give the extinction ratios.
The range of the magnitudes in each field provide an estimate of the
maximum extinction values. We find that the extinction in the bulge is
small and has a maximim value of about 0.25 in $E(B-V)$, 
in agreement with Paper I. 

The resulting values of the extinction ratios are listed in  Table
\ref{tbl:extinction} where the values are also compared with  the
Galactic values for the same $HST$ filters  based on the extinction
curve of Savage and Mathis (1979). The Galactic values were derived by
applying the extinction curve to the stellar energy distribution of an
unreddened A0-star and then convolving the resulting spectrum with the
$HST$ filter characteristics (for details, see Romaniello 1998, and
Romaniello \etal\ 2001). 
The table shows that the extinction ratios are
quite similar to those of our Galaxy, which suggests that the
extinction curve of the two galaxies are about the same.  Scuderi
\etal\ (2001) have reached the same conclusion for the M51 nuclear
region, but using a somewhat different method.

Hill et al. (1997) have shown on the basis of a study of H{\sc II}
regions observed with the {\it Ultraviolet Imaging Telescope}
that the far-UV extinction of M51 is smaller than in the Galaxy,
$A_{152}/E(B-V)=6.80$ for M51 compared to 8.33 for the Galaxy.
This is not in contradiction with our results, because the
far-UV extinction at $\lambda < 2100 \AA$ is known to vary 
drastically both in the Galaxy and in the LMC between different regions. 
 
%
%

%
\begin{table}
\caption[]{Reddening in the Galaxy and in the bulge of M51}
\label{tbl:extinction}
\begin{tabular}{lcc}
\hline
Ratio & Galaxy  & M51\\
\hline
                &   &   \\
$E(B-I)/E(B-V)$ & 2.22 & $2.14\pm 0.25$   \\
$E(V-I)/E(B-V)$ & 1.52 & $1.15\pm 0.25$   \\
$E(R-I)/E(B-V)$ & 0.66 & $0.37\pm 0.15$   \\
  &  & \\
$A_{UV}/E(B-V)$ &   6.71 & \\                  
$A_U/E(B-V)$    &   5.04 & \\
$A_B/E(B-V)$    &   4.17 & \\
$A_V/E(B-V)$    &   3.26 & \\
$A_R/E(B-V)$    &   2.46 & \\
$A_I/E(B-V)$    &   1.90 & \\
\hline
\end{tabular}
\end{table}

In the above described analysis we have treated the extinction as a
foreground effect. This is justified by the small values of $E(B-V)<0.25$,
for which the effect of the extinction on the flux is independent of
the depth distribution of the dust. The fact that we find an
extinction curve closely resembling the Galactic one provides additional
support to the justification of the adopted method.


\section{Bright point sources in the bulge}

The $HST-PC$ image of the central region of M51  (see Fig.
\ref{fig:bulge}) shows the presence of   bright pixel-size spots, with
$21.4 < V < 24.4$. 
 We have estimated the sizes of the sources, using radial
profile fitting using radial profile fitting of the F555W images. 
The vast majority, 21 out of 30, are definitely point sources.
Six sources have a too high background to allow a 
reliable size determination. All of them are rather blue with $B-V <
0.32$ (see below).
Two sources (nrs 11 and 13) are possibly extended, but they have a low
peak countrate and are rather faint. One of these, nr 11, is a very
red source with $B-V=1.8$ whereas nr 13 is a blue source 
with $B-V=-0.09$ (see below).
Only one source, nr 4 with $B-V=0.90$ is definitely resolved. 
If its intrinsic intensity 
distribution is Gaussian, it has an intrinsic 
FWHM of 0.06 arcsec, which corresponds to 2.4 pc at the distance of M51. 
Assuming that the sample of 
sources that have no reliable FWHM determination
contains the same fraction of extended sources as the rest, we
conclude that our sample may contain no more than 4 extended
sources.
From here-on the sources will be called ``bulge point sources''.

Could the bulge point sources be back-ground galaxies?
Using the number and magnitude distribution of galaxies in 
the Hubble Deep Field (Williams et al. 1996), Gonzalez et al. (1998)
derived the expected number and brightness distribution of
background galaxies in the $HST-WF$ chips of an observation similar to
the ones of M51. Our detection limit is about 23.5 magn in $V$ and
$I$, due to
the bright background by the old stellar background population in the
bulge. From the study of Gonzalez et al. (1998) we  find that
about 13 background galaxies in $V$ and 25 in $I$ above the detection
limit are expected
{in the three $WF$-chips together}. The bulge of M51 covers an area
wchich is only 17 \% of the $PC$-chip and 1.5 \% of the three
$WF$-chips together. So we expect that the bulge image contains
about 0.2 background galaxies in $V$ and 0.3 in $I$. Moreover
background galaxies would not be point sources.
Even the extended source nr 4, with its FWHM of 0.6 arsec, 
is too small to be a background galaxy (unless it is a QSO).
It is most likely a small cluster.

\subsection{The location of the bulge point sources}
 
The location of the bulge point sources is indicated in Fig.
\ref{fig:bulge}. The points are not randomly distributed, but seem to
be located preferentially  in an ellipse (with major axis running
from upper left to lower right) or in strings  around the nucleus.
This agrees with the general appearance of the region around the
nucleus as observed in the $B$ filter: the darker lanes also indicate
spiral arms which can be traced down to the nucleus. The main spiral
arms originate at a larger distance $\sim 13'' \sim 500 $ pc from the
nucleus, i.e. just outside the range of Fig \ref{fig:bulge}  (see
Figure \ref{fig:pcimage}).  

\begin{figure}
\centerline{\psfig{figure=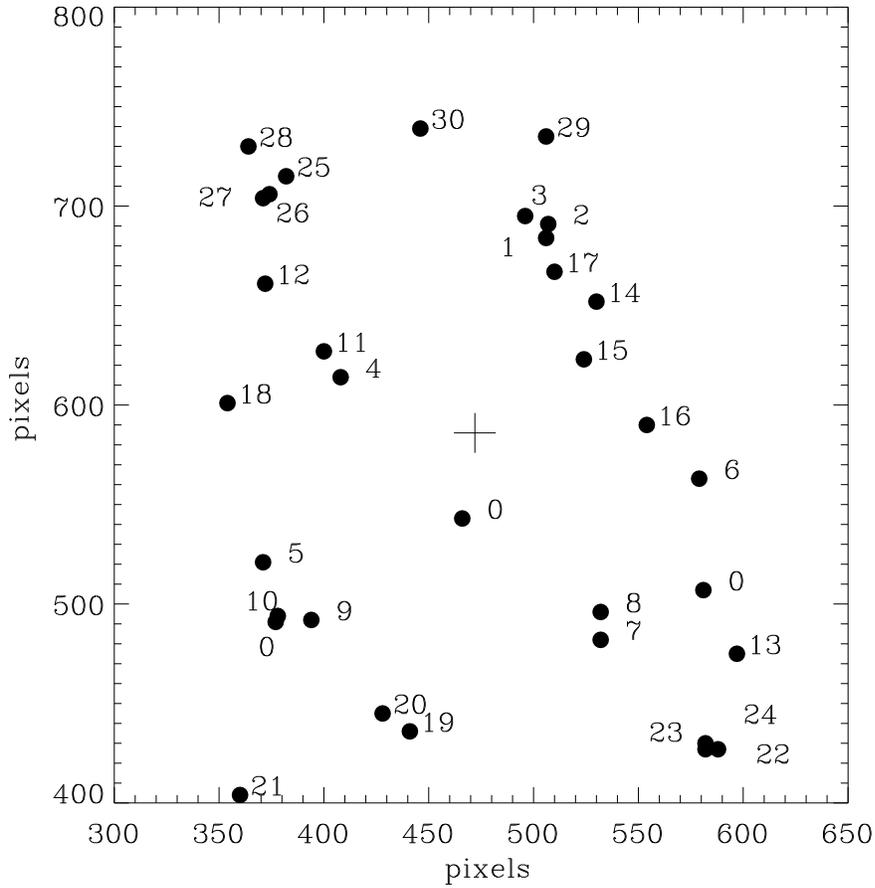}}
\caption[]{The location and the numbers of the 30 point sources in the
  Bulge of M51, that are studied in this paper. Sources with number
  zero were in the original selection (see Figure 2) but were not studied
  because of the lack of reliable photometry. The cross indicates the
  location of the nucleus of M51 (Paper I). The orientation is the
  same as in Fig. 2 }
\label{fig:bulgenrs}
\end{figure}

Figure \ref{fig:bulgenrs} shows the locations and the numberings of the
point sources. From the originally selected 33 sources, 3 have  been
removed because we could not derive reliable photometry. These are
indicated by the number zero. The numbers of the remaining 30 other
sources are indicated. The coordinates of the sources are listed in
Table 2..

%

\subsection{Photometry of the bulge point sources}

The integrated fluxes of the point sources were measured in the $UV$,
$U$, $B$, $V$, $R$ and $I$ bands, as described in Sect. 2, and
converted into Vega-magnitudes, where $UV = U = B = V = R = I = 0$ for
Vega (Whitmore 1995). Fourteen out of the 30 point sources could not be
measured in the $UV$ and $U$ band because they are too faint. For these
point sources we adopted safe brightness upper limits of $UV=21.4$ and
$U=22.3$,  which correspond to the magnitudes (minus its
uncertainty) of the faintest point sources that were detected in the
$UV$ and $U$ band images. Sources nr 26 and 27, which are located close
together,  could not be measured separately in the $UV$ and $U$ images
of the WF-chips. We measured the combined magnitudes of the two
sources: $UV=19.79 \pm 0.11$ and $U=20.51 \pm 0.06$. These two sources
are resolved in the other bands, where they are in the PC-chip. Since
the energy distributions of the two sources in the $B, V, R, I$ bands
are very similar with a mean magnitude difference of 0.30, we adopt the
same difference for the $UV$ and $U$ magnitudes, and derived the
magnitudes of the two individual sources. The magnitudes of all bulge
point sources  are listed in Table 2. 

Most sources
have $B-V$ colors between $-0.5$ and $+1.0$ with two red exceptions
near $B-V \simeq +2.0$  (nrs 3 and 11) and one very uncertain 
blue exception, namely
nr 16 with $B-V=-2.5 $. Most sources
with $B-V < +0.5$ are detected in the $UV$ and $U$ band. The three
bluest UV sources are nrs 12, 26 and 27 which have $-2.1 < U-V < -1.7$
and $-3.1 < UV-V < -2.6$. These must be hot objects with little or no
extinction. The colors of the majority of the point sources, in the
range of $ -0.3 \le B-V \le +1.0$, correspond to  spectral types
between O and G5, if there were no extinction. The reddest point
sources, nrs 3 and 11,  have $B-V=+2.2$ and +1.8 respectively, which
correspond to unreddened M stars.

%
\begin{table}
\centerline{\psfig{figure=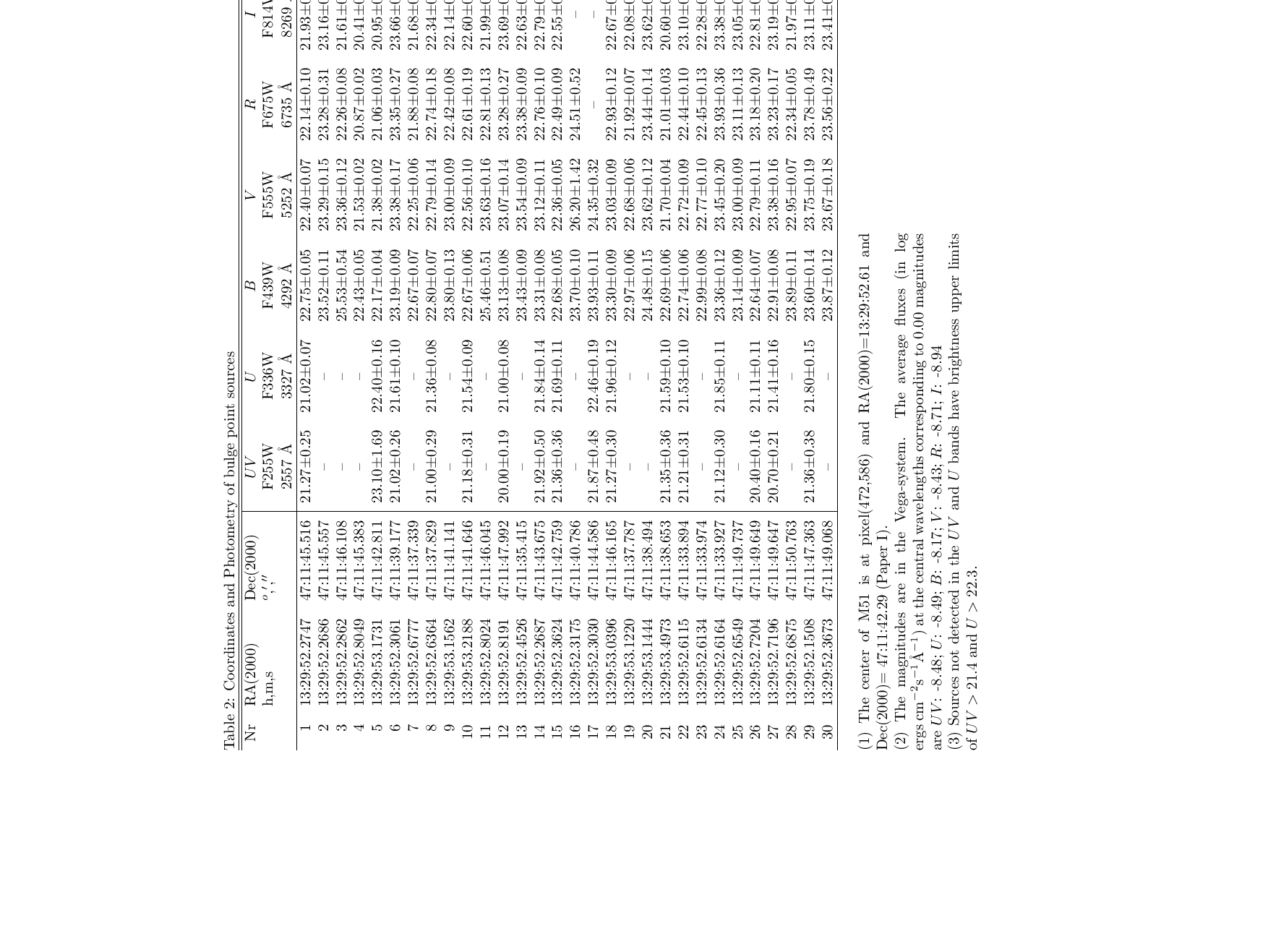,angle=90,width=24.0cm}}
\label{posphot}
\end{table}

\begin{table}
\caption[]{Bulge  point sources as clusters or stars}
\begin{tabular}{ll|lllr|lllr}
\hline
\hline
 &  & \multicolumn{4}{c|}{CLUSTERS} & \multicolumn{4}{c}{STARS}\\ 
Nr & nr obs& $E(B-V)$ & log($t$) & log($M_i$)$^a$ & $\chi_{R}^2$ &
$E(B-V)$ & log \Teff\ &  log \Lstar & $\chi_{R}^2$ \\
   &       & magn     &     yrs      &    \Msun    &          &  
 magn   &  K    &   \Lsun  &  \\
\hline
1     &6&  0.20 & 5.35$_{-0.05}^{+0.80}$ & 3.27     &  3.45 &
0.64 & 4.70$_{-0.22}^{+0.00}$ & 7.31$_{-0.68}^{+0.03}$ & 0.52\\
2     &4&  0.18 & 5.35$_{-0.00}^{+1.30}$ & 2.72     &  0.30 &
0.36 & 4.34$_{-0.44}^{+0.00}$ & 5.70$_{-0.58}^{+0.00}$ & 0.12\\ 
3     &4&  0.70 & 9.70$_{-0.80}^{+0.00}$ & 5.63$^F$ &  0.90 &
0.50 & 3.65$_{-0.00}^{+0.00}$ & 5.23$_{-0.04}^{+0.02}$ & 0.10\\
4     &4&  0.48 & 8.70$_{-0.22}^{+0.08}$ & 5.28$^F$ &  0.39 &
0.70 & 3.85$_{-0.00}^{+0.27}$ & 5.96$_{-0.00}^{+0.75}$ & 0.12\\
5     &6&  0.00 & 8.90$_{-0.06}^{+0.00}$ & 4.85$^F$ & 14.47 &
0.02 & 3.78$_{-0.00}^{+0.00}$ & 5.18$_{-0.02}^{+0.07}$ & 9.09\\
6     &6&  0.00 & 6.20$_{-0.00}^{+0.00}$ & 2.46     &  2.18 &
0.10 & 4.48$_{-0.10}^{+0.18}$ & 5.66$_{-0.32}^{+0.57}$ & 0.13\\
7     &4&  0.00 & 8.74$_{-0.09}^{+0.00}$ & 4.43$^F$ &  0.03 &
0.24 & 3.85$_{-0.00}^{+0.27}$ & 5.10$_{-0.02}^{+0.76}$ & 0.12\\
8     &6&  0.12 & 6.60$_{-0.00}^{+0.00}$ & 2.72     &  0.40 &
0.38 & 4.60$_{-0.26}^{+0.10}$ & 6.58$_{-0.76}^{+0.40}$ & 0.88\\
9     &4&  0.00 & 9.18$_{-0.18}^{+0.00}$ & 4.48$^F$ &  0.08 &
0.24 & 3.78$_{-0.04}^{+0.60}$ & 4.84$_{-0.22}^{+1.82}$ & 0.18\\
10    &6&  0.18 & 6.45$_{-0.10}^{+0.00}$ & 2.79     &  0.70 &
0.24 & 4.30$_{-0.00}^{+0.02}$ & 5.75$_{-0.03}^{+0.07}$ & 0.10\\
11    &4&  0.46 & 9.65$_{-1.48}^{+0.00}$ & 5.08$^F$ &  1.40 &
1.32 & 3.90$_{-0.33}^{+0.80}$ & 5.87$_{-1.07}^{+2.30}$ & 1.10\\
12    &6&  0.00 & 5.60$_{-0.00}^{+0.25}$ & 2.64     & 12.49 &
0.00 & 4.70$_{-0.00}^{+0.00}$ & 6.17$_{-0.00}^{+0.13}$ & 3.12\\
13    &4&  0.02 & 7.25$_{-0.00}^{+0.05}$ & 2.82     &  3.23 &
0.18 & 3.88$_{-0.03}^{+0.10}$ & 4.59$_{-0.21}^{+0.33}$ &30.18\\
14    &6&  0.28 & 6.20$_{-0.85}^{+0.20}$ & 2.88     &  1.58 &
0.54 & 4.70$_{-0.22}^{+0.05}$ & 6.65$_{-0.66}^{+0.20}$ & 0.51\\
15    &6&  0.02 & 6.65$_{-0.20}^{+0.00}$ & 2.76     &  5.00 &
0.34 & 4.30$_{-0.15}^{+0.40}$ & 5.93$_{-0.52}^{+1.15}$ & 3.19\\
16    &3&       &                        &          &       & 
     &                        &                        &     \\
17    &4&  0.00 & 7.00$_{-0.00}^{+0.00}$ & 2.71$^F$ &  6.15 &           
0.00 & 4.36$_{-0.00}^{+0.24}$ & 4.96$_{-0.00}^{+0.89}$ & 0.61\\
18    &6&  0.22 & 5.30$_{-0.30}^{+0.30}$ & 2.89     &  0.32 &
0.52 & 4.70$_{-0.30}^{+0.00}$ & 6.90$_{-0.90}^{+0.05}$ & 0.41\\
19    &4&  0.20 & 7.48$_{-0.00}^{+1.34}$ & 3.65$^F$ & 18.73 &
0.14 & 3.81$_{-0.00}^{+0.00}$ & 4.88$_{-0.05}^{+0.05}$ &20.54\\
20    &4&  0.00 & 8.85$_{-0.37}^{+0.00}$ & 3.88$^F$ &  5.63 &
0.00 & 3.78$_{-0.00}^{+0.00}$ & 4.22$_{-0.00}^{+0.10}$ & 4.28\\
21    &6&  0.18 & 9.18$_{-0.36}^{+0.37}$ & 5.22$^F$ &  2.86 &
1.22 & 4.65$_{-0.27}^{+0.05}$ & 8.17$_{-0.77}^{+0.16}$ & 0.63\\
22    &6&  0.20 & 6.45$_{-0.00}^{+0.00}$ & 2.79     &  1.89 &
0.30 & 4.38$_{-0.23}^{+0.27}$ & 5.97$_{-0.70}^{+0.86}$ & 1.76\\
23    &4&  0.20 & 6.50$_{-0.00}^{+0.00}$ & 2.95     &  0.17 &       
0.20 & 3.88$_{-0.03}^{+0.24}$ & 4.86$_{-0.14}^{+0.66}$ & 0.44\\
24    &6&  0.02 & 6.20$_{-0.05}^{+0.00}$ & 2.43     &  1.47 &
0.14 & 4.48$_{-0.13}^{+0.22}$ & 5.66$_{-0.48}^{+0.73}$ & 0.57\\
25    &4&  0.00 & 7.54$_{-0.00}^{+0.00}$ & 3.25$^F$ & 11.83 &
0.06 & 4.00$_{-0.05}^{+0.00}$ & 4.69$_{-0.14}^{+0.03}$ & 1.21\\
26    &6&  0.00 & 6.35$_{-0.00}^{+0.05}$ & 2.58     &  2.48 &
0.00 & 4.36$_{-0.00}^{+0.34}$ & 5.49$_{-0.00}^{+1.06}$ & 0.77\\
27    &6&  0.00 & 6.40$_{-0.00}^{+0.00}$ & 2.42     &  1.98 &
0.00 & 4.36$_{-0.00}^{+0.24}$ & 5.36$_{-0.00}^{+0.76}$ & 0.99\\
28    &4&  0.02 & 9.40$_{-0.22}^{+0.00}$ & 4.69$^F$ &  0.09 &
0.34 & 3.78$_{-0.00}^{+0.00}$ & 4.97$_{-0.00}^{+0.00}$ & 0.03\\
29    &6&  0.00 & 5.50$_{-0.00}^{+0.00}$ & 2.42     &  2.71 &
0.26 & 4.70$_{-0.30}^{+0.00}$ & 6.36$_{-0.98}^{+0.17}$ & 1.49\\
30    &4&  0.22 & 6.20$_{-0.55}^{+0.20}$ & 2.55     &  0.05 &
0.22 & 4.00$_{-0.12}^{+0.30}$ & 4.65$_{-0.33}^{+0.85}$ & 0.02\\
\hline\\
\end{tabular}

$^a$:  $M_i$ is the initial mass of the clusters.\\
$^F$:  Frascati cluster model was adopted.

\label{tbl:bulgestars}
\end{table}



\section{Modelling the energy distribution of the bulge point sources}

The bulge point sources have visual magnitudes between 21.4 and 24.4
corresponding to $-8.2 < M_V < -5.2$ if there were no extinction. For a
mean reddening of $E(B-V)\simeq 0.3$ (see below) the absolute
magnitudes are about $-9.1 \le M_V \le -6.1$.  This means that the
bulge point sources could be either small clusters or very bright
stars. We will consider both possibilities.

\subsection{Bulge point sources as clusters}

If the blue point sources are clusters, they cannot be very massive. For
instance a cluster with an initial mass of $10^6$ \Msun\  and an age of
$2 \times 10^6$ years will have a visual absolute magnitude of
$M_V=-15.3$ (Leitherer \& Heckman, 1995). At the distance of M51 and
with an extinction of $E(B-V)=0.3$ the cluster will have $V \simeq
15.3$. The observed point sources are 7 to 9 magnitudes fainter, so
their initial mass must have been a few $10^2$ to a few $10^3$ \Msun.
For clusters of such a small mass, the colors and magnitudes cannot be
accurately predicted because they will depend on the evolutionary stage
of one or few of the most luminous and most massive stars. Therefore, 
the results of fitting the observed energy distributions to models
should only be considered as a rough estimate of the cluster parameters.

 We have fitted the observed energy distributions of the bulge point 
sources with those predicted for instantaneously formed clusters,
from Leitherer \& Heckman (1995), henceforth called ``LH-models''
and from the Frascati-group (see below), henceforth called the
``Frascati-models''.

 We adopted the LH-models of clusters with solar metallicity,
with an IMF of $\alpha=2.35$ (Salpeter's value) and with an upper mass
cut-off of 100 \Msun\ and a  lower mass cut-off  of 1 \Msun. 
These models cover an age range from 0.2 to 300 Myrs.
The predicted magnitude in the $F255W$ band was derived from their
magnitude at 2100 \AA\ and the slope of the spectral energy
distribution between 2100 and 3000 \AA. 
The ''Frascati-models'' were calculated by Romaniello
(1997) from the evolutionary tracks of Brocato \& Castellani (1993) and
Cassisi et al. (1994) using the WFPC2 magnitudes derived from the
stellar atmosphere models by Kurucz (1993). 
These models are for instantaneous formation
of a cluster of solar metallicity stars in the mass range of 0.6 to 25
\Msun,
distributed according to Salpeter's IMF. These models cover an
age range of 10 to 5000 Myr. 
The LH-models are expected to more accurate for the younger
clusters, because they are based on the evolutionary tracks of the
Geneva-group which include massive stars. The Frascati models are
expected to be more accurate for the old clusters, because they
include stars with masses down to 0.6 \Msun.
For both sets of cluster-models the magnitudes were calculated 
directly from the predicted energy
distributions, using the $WFPC2$ filter calibration (Whitmore 1995). 
These magnitudes will be  compared directly with the observed magnitudes,
so we do not have to apply the ``Holtzman et al.'' (1995)
colour correction, which is needed when the magnitudes of the models
are given in the standard filters.

For fitting the predicted to the observed energy distributions we
used a three
dimensional maximum likelyhood method. The free parameters are
$E(B-V)$, the initial mass, $M_i$, and age, $t$. 
We corrected the observed magnitudes for extinction in the range of
$0.00 < E(B-V) < 2.0$ in steps of 0.02. The extinction values listed in
Table \ref{tbl:extinction} were adopted. The weighting factors are
chosen as  $w_i = 1/(\Delta m_i)^2$ where $\Delta m_i$ is the
uncertainty in the magnitude.  For sources that were not detected in
the $UV$ and $U$ images we adopted brightness upper limits of $UV > 21.4$
and $U > 22.3$, as discussed above.  The uncertainty in the
fitting of the $UV$ and $U$ magnitudes to the models depends more
strongly on the accuracy of the  adopted extinction curve than on the
accuracy of the observed magnitudes. This uncertainty of the short
wavelength fitting  will be larger for objects with large values of
$E(B-V)$.   To take this into account we have added an extra  term in
the uncertainty  of $\Delta UV = 1.5~E(B-V)$ and $\Delta U =
1.0~E(B-V)$  in the model fitting of the $UV$ and $U$ magnitudes, with
a minimum uncertainty of $(\Delta UV)_{\rm min} = 0.15$ and $(\Delta
U)_{\rm min}  = 0.10$.  This implies that the fitting had be done in
two steps: first with the normal values of $\Delta m_i$ to find the
best fit  of the cluster  parameters and $E(B-V)$ and then with the
extra uncertainty in the $UV$ and $U$ magnitudes. The values of
$E(B-V)$ of the second iteration turn out to be  very similar  to those
of the first iteration. 

 The resulting fits of the LH-models and the Frascati-models were
compared and one of the two was adopted, based on the following
criteria: (a) significantly smaller reduced $\chi_R^2$; 
(b) significantly smaller value of $E(B-V)$.
This last criterium is used because all reliable fits 
(small $\chi_R^2$) have a small extinction and because the study of
the extinction (see \S 3) showed that $E(B-V)$ is very small in the
bulge. So if the energy distribution can be fitted with a young
cluster of high extinction or an older cluster of low extinction,
we adopted the second solution.
The uncertainties in the ages and initial masses are
derived from the requirement that only fits with $\chi_R^2 <
\chi_R^2(min)+1$ are acceptable. This corresponds to the 68 \% confidence
range.

\begin{figure}
\centerline{\psfig{figure=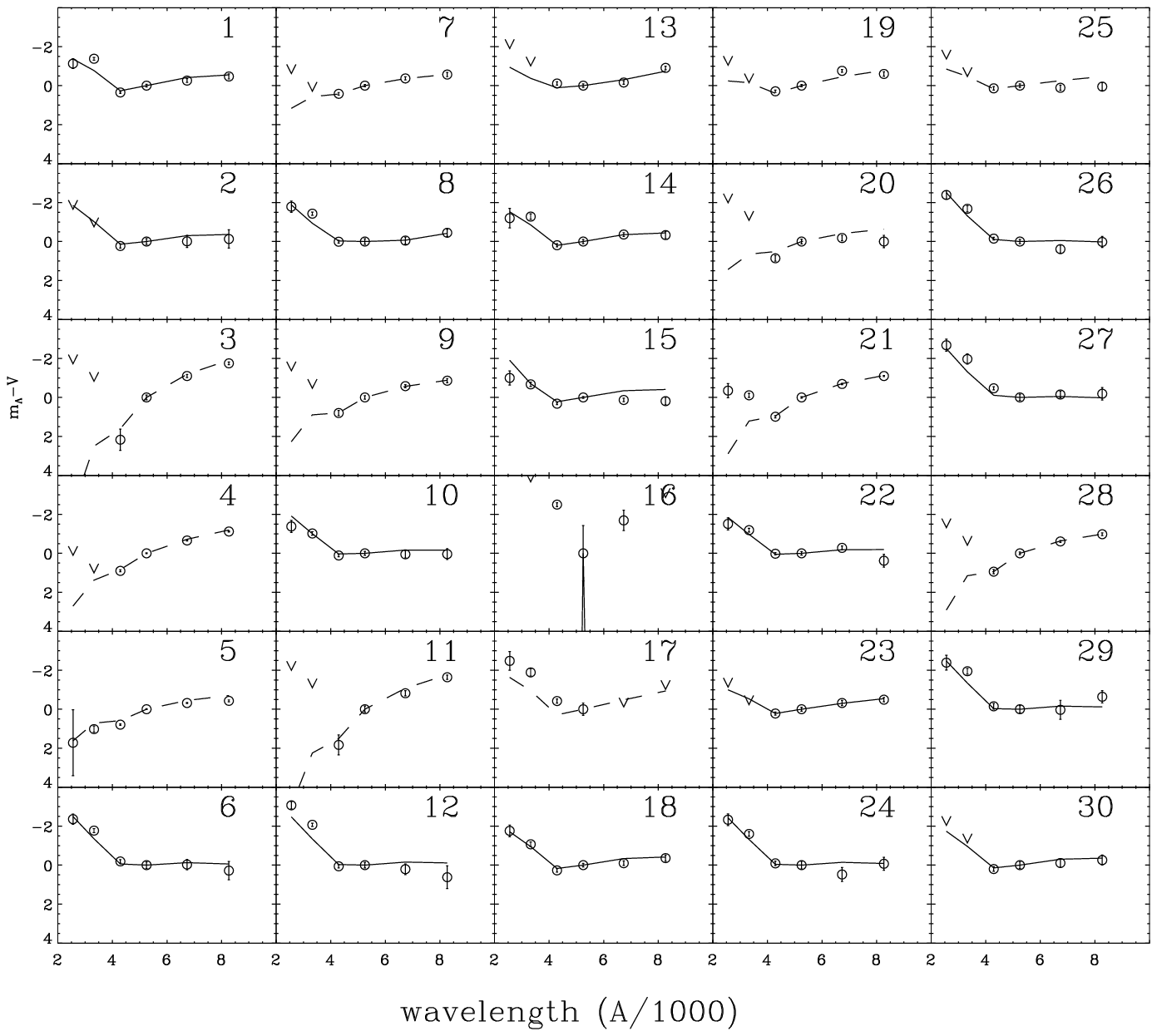,height=10.0cm}}
\centerline{\psfig{figure=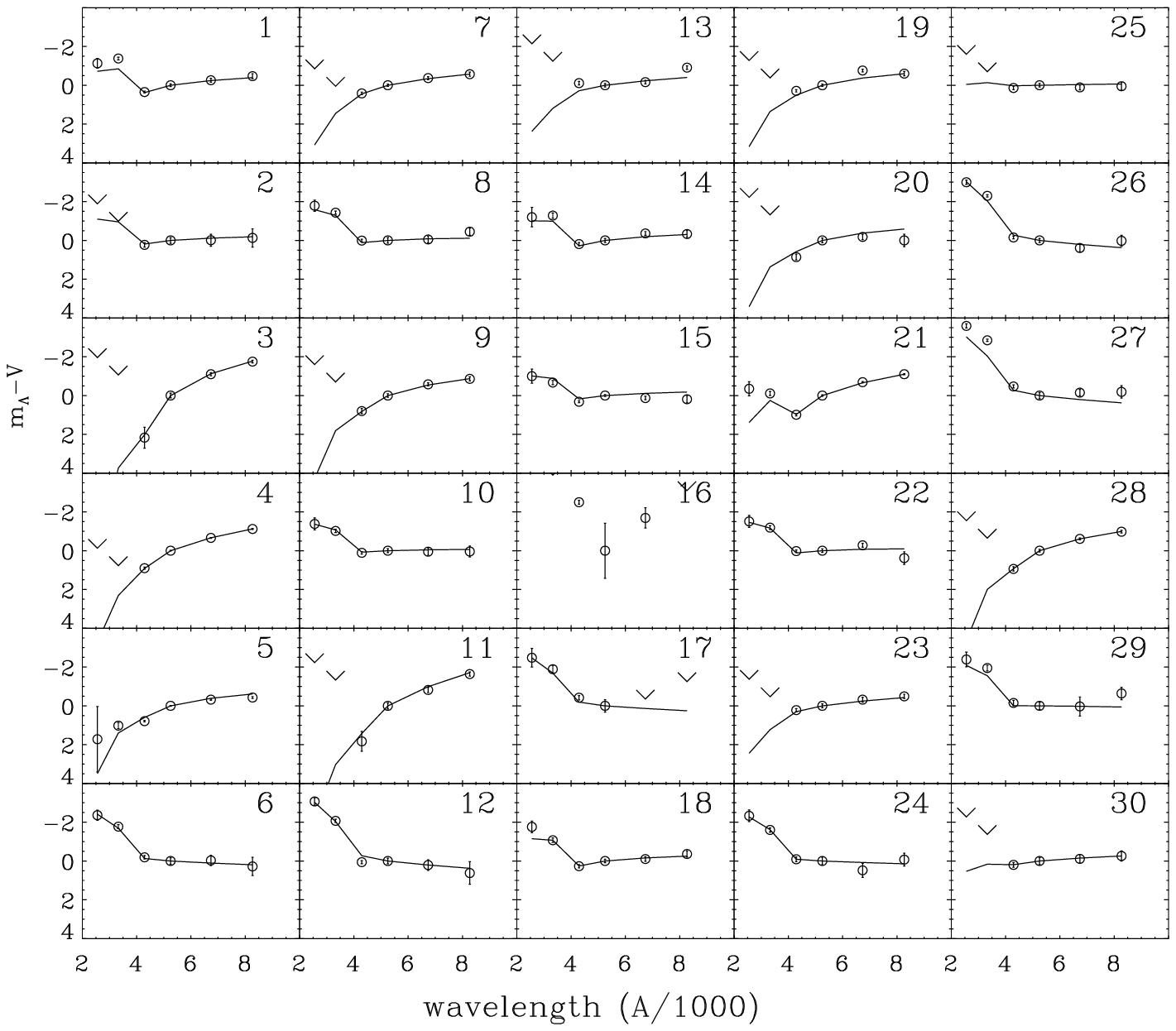,height=10.0cm}}
\caption[]{A comparison between the observed magnitudes of the bulge 
point sources (circles with errorbars) and the fits of cluster models
(upper figure) and stellar models (lower figure) with extinction. 
In the upper figure the full lines indicate Leitherer \& Heckman
(1995) models and dashed lines indicate ``Frascati-models''.
Arrows indicate upper limits.
 The vertical scale is
$m_{\lambda}-V$ in  magnitudes and the horizontal scale is
$\lambda(\AA)/1000$. In the upper figure the fits to 
LH-models are indicated by full lines, those of the Frascati models 
with dashed lines.
The fit  parameters are listed in Table 3.}
\label{fig:fits}
\end{figure}

 The results are listed in the left half of
Table \ref{tbl:bulgestars} and the fits are shown in Figure
\ref{fig:fits}a. For most sources a fit with a reasonably small
value of $\chi_R^2 \lesssim 3$ is found, however 7 
sources have a high value of $\chi_R^2 > 5$. The cluster parameters
of these models are not reliable.
 The results in Table \ref{tbl:bulgestars} confirm our previous
 estimate that the young clusters have a low mass. More than half of
 the clusters have an initial mass smaller than about 1000 \Msun.
Only old clusters with $t>500$ Myrs have $M_i>10^4$ \Msun.
This is due to the detection limit.
A cluster with an initial mass is below $5\times 10^3$
\Msun\ fades below the detection limit at an age of about
300 Myrs (Leitherer \& Heckmann, 1995).

\subsection{Bulge point sources as individual stars}

We have also fitted the energy distributions of the bulge point
sources   to those predicted by model atmospheres of Kurucz (1993).
For this purpose  we selected 15
models with temperatures ranging from 3750 K to 50000 K and the lowest
gravities in the grid of model atmospheres.  
 We realize that the Kurucz blanketed LTE-model atmospheres may not be  
very accurate for the hottest stars where Non-LTE effects and
atmospheric extension effects may play a role. So these models are not
expected to give accurate values of the stellar parameters based on $WFPC2$-photometry.
However, they
serve the present purpose of obtaining an indication of the effective 
temperature and luminosity of the bulge point sources.

The models were fitted to the observations using
a three dimensional maximum likelyhood method. The free parameters are
$E(B-V)$, $\Teff$ and radius \Rstar.
To fit the observed energy distributions to those of model atmospheres,
we corrected the observed magnitudes for extinction in the range of
$0.00 < E(B-V) < 2.0$ in steps of 0.02 (similar to the fit of the 
cluster models). 
The resulting dereddened energy distributions were compared to those of
the model atmospheres and the
best-fit model was determined.   The two fit parameters for the shape
of the energy distributions are  $E(B-V)$, \Teff.  The fit parameter
for the absolute magnitude is the angular diameter or radius of the
star. The weights of the fitting are the same as described in the 
previous section.  To take into account the uncertainty in the
extinction at the wavelengths of  the $UV$ and $U$ filter, we have
added an extra uncertainty $\Delta UV$ and $\Delta U$, as described in
the previous section.

 The results are listed in the right hand side of Table
\ref{tbl:bulgestars}. We list the values of $E(B-V)$, \Teff\ and log
$L/L_\odot$ and the value of the reduced $\chi_R^2$ of the fit.  
The range of acceptable parameters was determined in the same way as for
clusters, i.e. the acceptable models have $\chi_R^2 < \chi_R^2(min)+1$.
Notice
that for almost all sources which are observed in six wavelength bands,
the fits of the energy distributions to  stellar models are
significantly better (smaller $\chi_R^2$) than fits to the cluster
models. In total 23 objects have $\chi_R^2 < 3$ and only 3 have
$\chi_R^2>5$.

The values of $E(B-V)$ derived from the fits are in the range of  0.0
to 1.3, with more than half of the objects having $E(B-V) < 0.30$. The
uncertainty in the derived values of $E(B-V)$ affects the uncertainty
in \Teff\ and log $L$.  A higher value of  $E(B-V)$ corresponds to a
higher value of \Teff, a higher value of \Lstar\  (because the
bolometric correction increases with \Teff)  and a smaller radius of
the object.  We list the uncertainty in log \Teff\ and log $\Lstar$, 
but not in $E(B-V)$. The uncertainty in log $L_*$ is related to the
uncertainty in  log \Teff\ by   $\Delta {\rm log} L_* \simeq 6.7 \times
\Delta {\rm log} \Teff$ for the hottest stars with $\Teff > 30~000$ K
and $\Delta {\rm log} L_* \simeq 5.3 \times \Delta {\rm log} \Teff$ for
the stars with $10~000 < \Teff < 30~000$ K. For stars in the range of
$5 000 < \Teff < 10~000$ K, the BC is very small and almost independent
of \Teff. For the coolest point source in our sample, i.e. nr 3, the BC
is sensitive to \Teff\ but the uncertainty in the derived value of
\Teff\ is very small. Typical uncertainties are $\Delta E(B-V) \simeq
0.10$,  $\Delta$ log \Teff $\simeq$ 0.1 and $\Delta$ log \Lstar
$\simeq$ 0.4 for hot stars observed in  all six bands, and $\Delta
E(B-V) \simeq 0.20$, $\Delta$ log \Teff $\simeq$ 0.2 and $\Delta$ log
\Lstar $\simeq$ 0.6  for the cooler stars observed in the visual colors
only. For stars with $\Teff > 30~000$ K the uncertainty is larger
(possibly as large as 10~000 K at $\Teff > 40~000$) due to the
uncertainties in the adopted Kurucz (1993) model atmospheres.

Figure \ref{fig:fits}b shows the comparison between the observed 
magnitudes and the predicted magnitudes of the best fit model for each
point source  as a function of wavelength. The observed lower limits
are also indicated. Notice that the agreement is satisfactory  for
almost all models. We discuss some typical energy distributions.\\
-- Source 3 is the reddest object in our sample. It fits the energy
distribution of a cool star of \Teff = 4500 K  with a large extinction
of $E(B-V)=0.50$.\\
-- Sources 12, 26 and 27 are the three objects with the brightest 
relative $UV$ magnitude in our sample. They have an energy distribution
of a hot star of \Teff = 50~000 K (nr 12) or 23~000 K (nrs 26 and 27)
without extinction.\\ 
-- Source 20 is a cold star of \Teff = 6000 K without extinction.  The
sharp decrease in flux to shorter wavelength agrees with the observed
upper limit  of the flux in the $UV$ and $U$ band.\\
-- If source 21 is a hot but heavily extincted star, as suggested by
the fit, its luminosity is so high that it must be a cluster.
The cluster fitting showed that it could be an old cluster.
Surprisingly, however, the FWHM of this source shows that it is a 
point source. Maybe it is a very cool star. \\

 From studies of Galactic early type  stars  it is known that about
half of the  luminous stars are in binary systems with a mass ratio
close to unity (Garmany \etal\ 1980).  So we can expect that a significant
fraction of the bulge point sources are in fact binaries. 
Since the post-mainsequence lifetime is much shorter than the main
sequence lifetime it is most likely that the systems are observed
either when both stars are on the main sequence, or when one of the
two stars has already finished its life. 
So with a few exceptions (less than about 10 \%), 
we can expect that the energy distribution of
a binary system will be close to that of a single star with a
luminosity of at most twice as high as that a single star.

\begin{figure}
\centerline{\psfig{figure=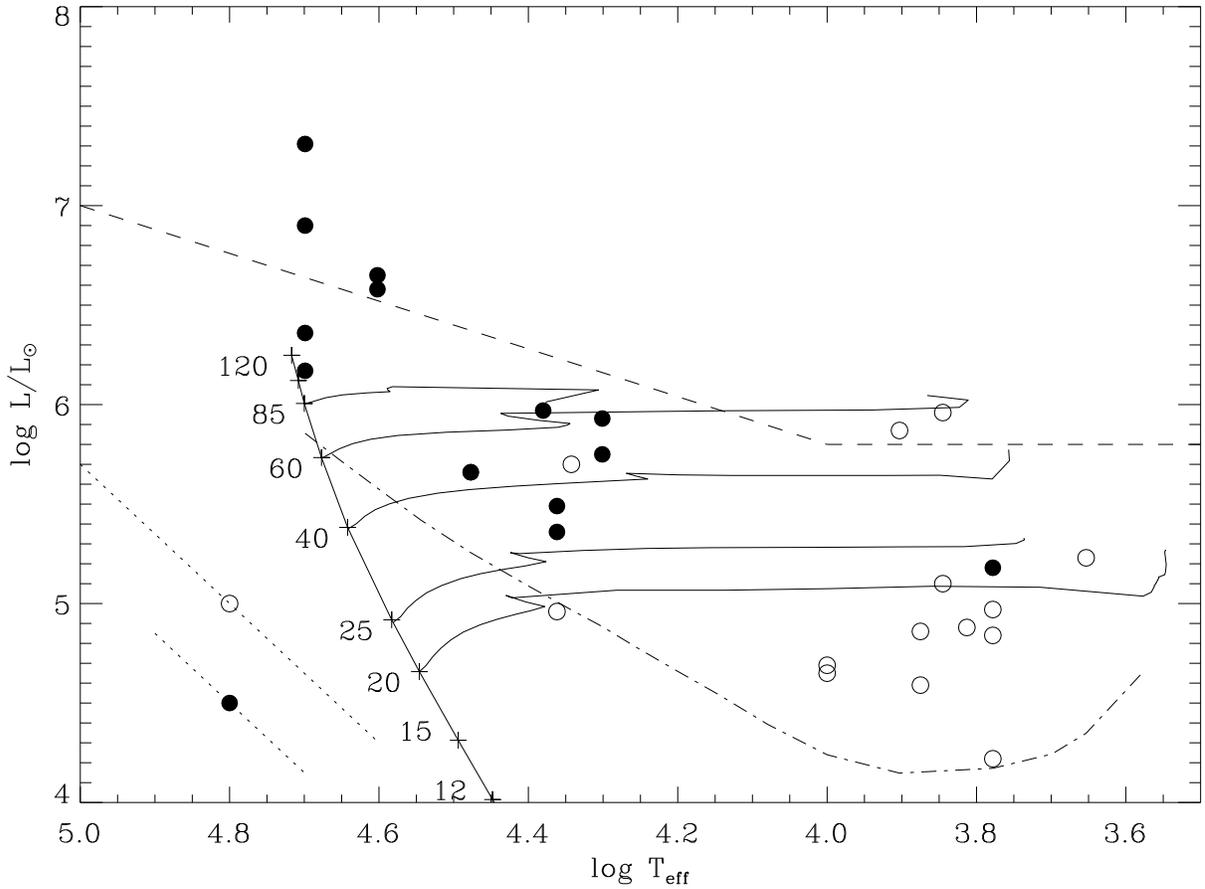}}
\caption[]{The HR diagram with the point sources in the bulge of M51,
under the assumption that they are single stars.   Open circles refer
to stars that were not detected  in the $F255W$ and the $F336W$ images.
Typical errorbars are indicated. The full lines are the evolutionary
tracks for Galactic stars with enhanced mass loss rates from Meynet et
al. (1994). The main sequence is indicated with initial masses. The
dashed line is the Humphreys-Davidson luminosity upper limit for
Galactic stars. The dash-dotted line is the predicted detection lower
limit for unreddened stars of $V=24.0$. }
\label{fig:bulgehrd}
\end{figure}

Figure \ref{fig:bulgehrd} shows the resulting HR diagram of the bulge
point sources. One object, nr 21 with $\log \Teff = 4.65$ and
$\log \Lstar/\Lsun=8.17$, is outside the range of the figure. An
uncertainty in $E(B-V)$ and correspondingly in \Teff\ and  log \Lstar\ 
for a fixed value of $V$ results in tilted error bars, because a higher
$E(B-V)$ implies a higher temperature and a larger bolometric
correction.  We show only the typical error bars as the individual
values can be derived from Table \ref{tbl:bulgestars}. The figure also
shows part of the evolutionary tracks of massive stars for Galactic
metallicities from Meynet et al. (1994) for enhanced mass loss rates. 
We see that {\it if} the point sources are indeed single stars, their
initial masses are in the range of about 12 to 150 \Msun. The four
brightest objects might even have masses of about 200 \Msun, if they
are single stars. This is about the same mass as the famous ``Pistol
star'' near the Galactic center (Figer et al. 1998).

The empirical luminosity upper limit for stars, i.e. the Humpreys-Davidson
limit, HD-linit, (Humphreys \& Davidson 1979; 
Fitzpatrick \& Garmany 1990), is shown in Figure \ref{fig:bulgehrd}.
Almost all the bright point sources have luminosities very close to or 
below the HD-limit. It suggests that at least a majority of the bulge
point sources could be massive stars. The group of hot stars with $\log
\Teff \ge 4.30$ and $\log \Lstar/\Lsun \ge 5.5$ have initial masses
between about 40 and 150 \Msun\ and ages on the order of 4 to
$8\times 10^6$ years or less. The group of cool stars with $\log \Teff
\le 4.0$ and  $\log \Lstar/\Lsun \le 5.3$ have initial masses between
15 and 25 \Msun\ and ages between 7 to 17 $\times 10^6$ years (Meynet
et al. 1994).
 We note that the radius (or luminosity) of the star was a free
  parameter in the fitting of the models to the observations, so the
luminosity was not restraint by the input models.

The two hot objects high above the HD-limit in Fig. \ref{fig:bulgehrd}
are:\\
-- nr 1 with $4.48 < {\rm log}~ \Teff < 4.70$, 
and $6.63 < {\rm log}~ \Lstar /\Lsun < 7.34$ and\\
-- nr 14 with $4.38 < {\rm log}~ \Teff < 4.65$, 
and $5.99< {\rm log}~ \Lstar/\Lsun < 6.85$.\\
Nr 1 is definitely a point source. Nr 14 is in a region of high
background radiation, so its FWHM could not be determined. 
These sources might be either  superluminous stars or small clusters.
Source 21 is also a cluster, based on its brightness.
 We note that due to the uncertainties in the
Kurucz model atmospheres of high temperature, the hottest objects may have 
a substantial uncertainty
in the determination of $\Teff$ and $\Lstar$.

The lower limit of the luminosity  of the point sources in the HR
diagram is due to a combination of three effects: (a) the detection
limit, (b) the extinction and (c) the Bolometric Correction as a
function of intrinsic color. We can predict the location of the lower
limit in this diagram if we adopt a magnitude limit of $V<V_{\rm
lim}\simeq 24.0$, no extinction and the BC versus \Teff\ relation from
the Kurucz (1993) model  atmospheres for Galactic metallicity. The
resulting lower limit is shown in Figure \ref{fig:bulgehrd}. It agrees
very well with the observations. Stars (or clusters) fainter that this
detection limit might be present in the bulge of M51, but would not
have been detected.

A striking feature of the distribution of the stars in the HRD is the
gap in temperature between $\log \Teff = 4.0$ and 4.3, apart from the
errorbars of the sources just outside the gap. 
 The mean value of $\log(\Teff)$ is $4.51 \pm 0.16$ for the hot group
of $\log(\Teff) >4.2$ and $3.84 \pm 0.09$ for the cold group of
$\log(\Teff) < 4.0$, so the two groups do not overlap.
Such a gap in the \Teff -distribution is expected
if the point sources are stars, because massive stars cross this
temperature interval in a short time after their main sequence phase.
The gap is not expected if the sources were clusters.

\subsection{The contribution of low mass stars to the spectral energy
distribution of point sources}

We found above that the energy distributions of the majority of the 
sources can best be fitted to that of individual massive stars,
rather than clusters. Moreover, the vast majority of the objects are
point sources (see \S 4). This shows that the energy distributions
of the objects is dominated by one (or very few) massive star(s).
In this section we will 
try to answer the question: `` how many low mass stars could be
hiding near the luminous stars before we would notice their presence
in the spectral energy distribution?''. 
We will describe two tests.

\subsubsection{Test 1: a cluster with an IMF mass distribution}
\label{sec:test1}

Let us consider the case that an observed 
point source is in fact a cluster with a certain
IMF.  Suppose that the cluster contains one massive star (the observed
one), with the derived mass $M_*$ and  luminosity $L_*$, plus a tail of
lower mass stars on the main sequence, distributed in mass according to
an IMF  $N(M)=C~M^{-\alpha}$ with $\alpha= 2.35$ (Salpeter's value). In
these calculations we  adopt a conservative lower mass limit of
$M_{\rm min}= 1 \Msun$. The constant C is determined from the condition
that the most massive star is formed at the median of its probability
interval, so that

\begin{equation}
\label{eq:median}
\int^{M_{\rm max}}_{M_*} N(M)dM~=~
\frac{C}{-\alpha +1}\left(M_*^{-\alpha+1}-M_{\rm max}^{-\alpha+1}\right)~=~0.5
\end{equation}
where $M_{\rm max}$ is the mass upper limit for star formation, which
we assume to be about 200 \Msun. This will then also
determine the mass of the next massive star, as the solution of the
same integral but with $M_*$ replaced by $M_{\rm next}$ and
0.5 by 1.5, etc. 
If we also assume a mass--luminosity relation $L/\Lsun \propto
(M/\Msun)^{\beta}$ with $\beta=3$, we can calculate the total
luminosity and the total mass of the cluster.  

We found that in a cluster with a most massive star of
$M_*=80$ \Msun\ the stars with masses $M<80 \Msun$
contribute only about 1 percent to the luminosity when  their total mass is
$640 ~\Msun$.  For a cluster with a most massive star of 120 \Msun\ 
a total mass of 1100 \Msun\ in lower mass stars is needed to
increase the luminosity by 1 percent.
 This contribution from
lower mass stars scales linearly with the mass of the cluster.  The
mean colors of the total radiation from cluster stars of $M_{\rm
min}<M< \Mstar$  are similar to those of an A type star, i.e. $BC \simeq
B-V\simeq V-R \simeq 0$.

The next question is: can we notice the presence of such a cluster in
the energy distribution? To answer this question  we have to look at
the long wavelength part of the spectrum,  where the contribution by
the cool stars may dominate that of the brightest hot star.  The most
stringent test is provided by the bluest point sources. The bluest
source without extinction is nr 12, which has $\Teff=50~000$ K, $\log
\Lstar=6.17$, $M_* \simeq 120~\Msun$ and a bolometric correction of
4.27. The visual magnitude has an uncertainty of 0.14 (1 $\sigma$). If
there is a cluster, it should  increase the brightness in the $V$-band
by less than $3 \sigma (V)=0.42$. Suppose the brightest star (i.e. the
observed hot star) has a luminosity of $L_*$ and the contribution of
the cluster to the luminosity is  $x L_*$. It is easy to show that the
visual brightness will increase by a factor

\begin{equation}
 \frac{F_V(\rm bright~ star~ plus~ cluster)}{F_V(\rm bright~ star)}~=~
1 + x~10^{-0.4 BC(\rm cluster)+0.4 BC(\rm star)}
\label{eq:Vcluster}
\end{equation}
where the two bolometric corrections are those of the brightest star
and of the rest of the cluster. The factor $x$ is thus related to the
brightness increase in the $V$-band by

\begin{equation}
x ~=~ \frac{10^{+0.4 \Delta(V)}}{10^{-0.4 BC(\rm cluster)+0.4 BC(\rm star)}}
\label{eq:xcluster}
\end{equation}
with $\Delta(V)=3 \times \sigma(V)$. Applying the equation to star nr
12 we find $x=0.09$. Using the calculations described above we find an
upper limit to the  cluster mass of
840 \Msun. Using the same method to stars nrs 29, 6, 27 and 26, with
\Teff\ = 50 000, 30 000, 23 000 and 23 000 respectively, we find upper
limits to the mass of the clusters of 1200, 500, 420 and 240 \Msun.  
These values are of about the same order as derived from the cluster
fits in \S~5.1.
So
stars nrs 26 and 27 with $\Mstar \simeq 35 ~\Msun$ provide the most
stringent upper limits to the mass of the clusters, of only a few
hundred \Msun. A similar test for the $R$-band magnitudes gives almost
identical results.

\subsubsection{Test 2: a cluster with stars of $1 < \Mstar < 10 ~\Msun$}
\label{sec:test2}

In the second test we have assumed that the luminous star is surrounded
by a cluster consisting of stars with masses between 10 and 1 \Msun,
distributed according to an IMF with a slope of 2.35. We assumed that
the stars are on the zero age main sequence, and we adopted the
$HST$-band fluxes calculated by Romaniello (1999) for the Kurucz (1993)
model atmospheres to calculate the $HST$-magnitudes of such a cluster.
For a total cluster mass of 1000 \Msun,  at a distance of 8.4 Mpc we
find the following magnitudes (in the Vega system): $m_{\rm
bol}=22.12$, $UV=22.24$, $U=22.76$, $B=23.70$, $V=23.78$, $R=23.82$ and
$I=23.86$. We then fitted the observed energy distributions of those 14
point sources that were observed in all 6 wavelength bands, with a
model energy distribution that had the following properties: it
consists of a star with either \Teff=40~000 or 25~000 K (depending on
the value of \Teff\ that we found by assuming it to be a single star,
see Table \ref{tbl:bulgestars}), with a luminosity of $L_*$ plus a
cluster with a total mass of $M_{\rm cl}$. The free parameters of the
fitting procedure are: $L_*$, $M_{\rm cl}$ and $E(B-V)$. The fitting
was done in exactly the same way as described in Sections 5.1 and
5.2.  

For 9 out of the 14 point sources we find that the cluster masses of
the best fitting  energy distribution is less than 400 \Msun.
The energy distributions of sources
6, 12, 14, 15 and 29 are best fitted without clusters ($M_{\rm cl} \le
20~ \Msun$). For 6 out of the 14 sources (nrs 6,  10, 15, 24, 26 and
27) even the {\it maximum} cluster mass  that
is compatible with the observed energy distribution is less than 
400 \Msun.  

So we conclude that for the majority of the bulge point sources the
energy distribution is dominated by only one (or very few) massive
star(s), and that for the hottest sources the upperlimit to the 
mass of a possible accompanying cluster is surprisingly small.

\subsubsection{Comparison with the Orion Nebula Cluster}
\label{sec:test3}

We compare the energy distribution of the  hottest and
bluest sources of M51 with that expected for the very young Orion
Nebula Cluster (ONC), if that was at a distance of M51.

In optical and UV images, the ONC cluster with an age of about $3
\times 10^5$ yrs, appears like a group of four bright hot stars with a
distribution tail of much fainter and cooler pre-main sequence stars.
At the distance of M51 such a group of a few luminous hot stars might
ressemble the bulge point sources. In reality however, the Trapezium
stars are members of a larger star forming cluster  with a diameter of
about 5 pc, with about 1600 optical stars and a total mass in excess of
$10^3$ \Msun\ (Hillenbrand, 1997). The main sequence is populated down
to $M_V \simeq +2$, corresponding to  type A5 and a mass of 2 \Msun.
The lower mass stars are still in their pre-main sequence phase. A large
fraction of the stars are  reddened by an extinction of $A_V= 1.5$ to
8.5  magnitudes.  
This is much higher than the extinction of the M51
bulge point sources. So the total energy distribution of the ONC stars
does not resemble that of the UV-brightest bulge point sources. 

At an age of a few Myrs, when most of the dust
around the ONC stars will be  dispersed and the lower mass stars have
reached the main sequence, the Orion nebula cluster would  look like a
``normal'' cluster with a well populated main sequence, weighted
towards the low mass end. We compare the energy distribution of the
unreddened ONC  with that of the hottest bulge point sources.

 Hillenbrand (1997) has determined the stellar parameters of 940 stars
of the 1600  ONC stars. She has shown  that this is a representative sample of the
stars in that cluster. The total mass of these 940 stars is 640 \Msun :
about 90 \Msun\ for the four hot Trapezium stars and 550 \Msun\ for the 
lower mass stars. These lower mass stars are under-represented by a
factor two.
We have used the parameters of these 940 stars,
together with the energy distributions of the stellar atmospheres
(Kurucz, 1993) to calculate the energy distribution of the cluster if
it was located in M51. The $V$-magnitude of this cluster, in the
HST-Vega system used throughout this paper,  would be 
$23.26$ if the stars had no reddening at all.
The four hottest stars contribute 84, 70 and 58 percent  to the
flux in the $UV$, $V$ and $I$ band repectively.
A comparison of the resulting unreddened ONC energy distribution with those
of the hottest M51 point sources shows that the spectra of
the bluest sources nrs 6, 24, 26, 27 and 29 
are compatible (within the uncertainty of the observations) with that
of the ONC.  The visual magnitudes of these point sources
are about the same as expected for the unreddened ONC. 
So these blue point sources look like the ureddened ONC if the low
mass stars are underrepresented by a factor 2. If we correct the
predicted ONC energy distribution for the
factor two 
undersampling of the low mass stars, we find that the predicted $R$
and $I$ magnitudes are too bright compared to the observed energy
distribution of the hottest bulge point sources.
Source 12 is ``bluer'' than the unredenned ONC, even with the undersampling
of the low mass stars.

We conclude from these tests that the M51 bulge point sources 
{\it could} be hiding small clusters. However for the bluest sources
these clusters must have a total mass of less than about 400 $\Msun$
whereas the star that dominates the energy distribution is already
more massive than about 40 \Msun\ or even 120 \Msun\ (nr 12).


\subsection{Stars or clusters?}

We have found that the energy distributions of the bulge point sources
can be fitted with those predicted for young massive stars or 
for young clusters of low total mass.\\
-- If they are single stars, the initial mass of the hot luminous  ones
with  $\log \Teff > 4.3$ and $\log \Lstar/\Lsun > 5.5$ must be larger
than about 40 \Msun, and their age must be lower than 4 Myrs. The group
of stars at $\log \Teff < 4.0$ and $4.2 < \log \Lstar/\Lsun < 5.2$  had
initial masses between 12 and 25 \Msun\ and an age of 7 to 17 Myrs.\\
 -- If the bulge point sources are clusters, half of 
the clusters  have ages larger than 10 Myr.
The other half of the clusters is very young ($<$ 4 Myr) with small 
initial masses between 250  and 2000 \Msun. The spectral energy
distributions of these young clusters show 
that even the clusters with the lowest masses must contain at least 
one massive hot star.

 There are four reasons suggesting that the point sources 
 are massive   stars or very
small groups of a few massive stars, or poor clusters 
whose energy distributions are dominated by one or two massive stars.

\begin{enumerate}
\item The single star models fit the energy distributions
systematically better than the cluster models.  This can be seen by
comparing the values of $\chi_R^2(min)$ of the stellar fit and the cluster
fit in Table \ref{tbl:bulgestars} and from Figure \ref{fig:fits}.
Basically, the observed fluxes of the point sources better fit the
``narrow'' energy distributions of single stars than the ``wider''
distributions predicted for clusters that contain a range of stars of
different masses and temperatures.
\item Half of the objects are very hot and should contain  at least one
or several massive O-stars in the range of 40 to 120 \Msun\ to explain
the UV magnitudes.  The tests described above show that {\it if} the
point sources are clusters, the cluster mass in the form of lower mass
stars is less than about  400 \Msun\ for the six $UV$ brightest
objects.
\item The point sources follow the Humpreys-Davidson luminosity  upper
limit in the HR diagram, with a few exceptions.  This is to be expected
in case the point sources are stars, but there is no  reason why
clusters would follow this stellar upper limit in the HR diagram.
\item The location of the sources in the HR diagram  shows a division
of the temperature distribution into two separate groups: one with
$\log \Teff > 4.30$ and one with $\log \Teff < 4.0$. This is to be
expected if the point sources are stars, because the temperature range
between the main sequence and the red supergiants in the HR diagram is
crossed rapidly by stellar evolution, but not if the point sources were
clusters. 
\end{enumerate}

 We realize that the number of sources is small and that we cannot fully
exclude the possibility that arguments 3 and 4 are the result of 
a chance coincidence. However the combination of the four arguments
strongly suggests that most of the bulge point sources are massive
stars or small groups of a few massive stars or poor
clusters whose energy distribution is strongly dominated by one or two
massive stars.
These are statistical arguments, based on the whole
sample of the bulge point sources. It is certainly possible that some
of the point sources are in fact clusters, instead of single or very
small groups of stars. This is probably the case for the few most
luminous hot objects with $\Lstar > 10^7$ \Lsun\ and certainly for
object 21 with $\Lstar > 10^8$ \Lsun .

\section{The star formation rate in the Bulge of M51}

If the point sources are single stars or small groups of stars, 
their age is very small and on the order of a few $10^6$
years. Source nr 13 with $t \simeq 2 \times 10^7$ years is the only
exception. (Even if they are clusters, at least half of them are very young
with ages leass than $4 \times 10^6$ yrs.)
This clearly indicates  the presence of ongoing  formation
of massive stars in the bulge. This is most likely related to the
morphological structure of the bulge. The old background population of
the bulge shows a smooth distribution of stars older than 5 Gyrs (Paper
I). However the dust distribution shows evidence for spiral-like
dust lanes. Figure \ref{fig:pcimage} shows that these dust lanes in the
bulge follow the same pattern and are the inner extensions of the
dust lanes observed outside the bulge.  One of the major dust lanes in
the bulge can be traced down into the North side of the nucleus and the
other one can be seen to enter the nucleus on the South side. The bulge
point sources also show evidence of occurring in strings  with a
morphology similar to that of the dust lanes. 
This shows that star formation is still going on in and near the
spiral-like dust lanes in the bulge of M51.

The star formation rate in the bulge can be estimated only in a rough
way. For this we consider the group in the HR diagram  of hot
stars with $\log \Lstar/\Lsun >5.5$ and the group of cooler stars with
$\log \Lstar/\Lsun < 5.2$ separately. The group of point sources with 
$\log \Lstar /\Lsun > 5.5$ contains stars of
initial mass in excess of 40 \Msun. Their main sequence lifetime is
about $4 \times 10^6$ years, depending only weakly on luminosity.
With an age of $4 \times 10^6$ years and a total mass of about 1400
\Msun\ (Fig. \ref{fig:bulgehrd}), the  formation rate of these most
massive stars in the bulge is $3 \times 10^{-4}$ \Msunyr. 

We can also estimate the star formation rate from the 13 fainter point
sources of $\log \Lstar/\Lsun < 5.2$, corresponding to stars in the
initial mass range of 12 to 20 \Msun. In this mass range we do not see
the stars in the main sequence phase because they will be too faint
(see the lower limit in Figure \ref{fig:bulgehrd}), but instead we see
them in the red supergiant phase. The red supergiant phase of stars in
the range 12 to 20 \Msun\ lasts about $1.2 \times 10^6$ years (1.6 Myrs
for 12 \Msun\  and 0.7 Myrs for 20 \Msun ).  The observed sources have
a total initial mass of about 200 \Msun. With this mass and  an age
range of $1.2 \times 10^6$ years we estimate the star formation rate of these
lower mass stars to be  about $2 \times 10^{-4}$ \Msunyr, i.e. of the
same order of magnitude as found for the most massive stars. Taking
both groups together, we find that in the mass range of about 12 to 120
\Msun\ the minimum star formation rate in the bulge is about  $5 \times
10^{-4}$ \Msunyear.  (In this estimate we did not include the mass of
source 21, which is most likely an old cluster).
This estimate provides obviously only a lower limit because 
we have assumed that only the most massive stars with  $M > 12 \Msun$
are formed. In reality stars may be formed over a whole mass range down
to some lower mass limit. This effect can be taken into account by
assuming a continuous star formation with a given slope of the IMF:
$-\Gamma \propto d(\log~N)/(d\log~m)$ where $N$ is the number of stars
per unit logarithmic mass interval. 
If we adopt $\Gamma=1.35$ (Salpeter's value) 
with a lower limit of 1 \Msun, we find a 
star formation rate of $2 \times 10^{-3}$ \Msunyr. If we adopt a
flatter IMF of $\Gamma=0.65$ we find a rate of $7 \times 10^{-4}$ \Msunyr.

We can derive a maximum star formation rate by assuming that all the bulge
point sources are clusters, and using the cluster masses and ages
listed in Table \ref{tbl:bulgestars}. Taking the masses of the
clusters formed in the last 10 Myrs, we find a total mass of $1.6
\times 10^4$ \Msun\ and a formation rate of $1.6 \times 10^{-3}$
\Msunyr. This is very similar to the rate derived under the assumption
that all the point sources are young stars. This similarity is due to the very
small masses of the clusters, compared to the high mass of the stars, 
plus the correction for the presence of lower mass stars.

In Paper I we found that the bulge has a total dust mass of $2.3 \times
10^3$ \Msun.  The gas content of the bulge in the form of neutral H has
been  derived from the 21 cm observations by Tilanus \& Allen (1991).
They find that the column density of neutral H in the bulge is much
smaller than outside the bulge. In the inner 1 arcminute, i.e. within
240 pc from the nucleus, they measured  a column density of $2 \times
10^{20}$ H atoms~cm$^{-2}$, with a maximum of $4~\times 10^{20}$ 
(Allen, Private Communications). This implies a total HI mass of $4
\times 10^5$ \Msun\ and a gas-to-dust ratio of about 170, which is very
close to the mean value of $140 \pm 50$ in the inner parts of our
Galaxy (Cox 2000, p 160). The estimate of the gas content
excludes  the contribution of the regions of high CO density within 4
arcsec = 160 pc  from the nucleus found by Scoville et al. (1998) which
have a total mass of about $10^7$ \Msun.) 

Adopting a total (gas + dust) content of $4 \times 10^5$ \Msun\ and a
star formation rate of 1 to 2 $\times 10^{-3}$ \Msunyr, we find that
the bulge could sustain this star formation rate during about only 2 to
$4 \times 10^8$ years. Interestingly, this is also the age of the starburst
in the nucleus of M51 and the approximately the time 
of closest approach of the
companion (Paper I). This strongly suggests that the massive star
formation  in the bulge that we see now is fed by material that was
brought into the bulge by the interaction of the two galaxies.


\section{Discussion}

The most surprising result of this study is the presence of bright point
sources, in the bulge of M51 that is otherwise  dominated by an old
background stellar population and spiral-like  dust bands of moderate
 optical depth ($E(B-V)\simeq 0.2$). The energy distribution of the
point sources shows that they are most likely isolated (or very small
groups of) massive stars, or very small clusters which are 
completely dominated by
one or few massive stars. The distribution of the point sources in the
Bulge of M51 in
``strings'' (see Fig. \ref{fig:bulgenrs}) shows that they are not
ejected from the starburst in the nucleus, but must have been formed 
in situ.
This shows that massive stars do not
necessarily form in clusters but that they can be formed as isolated
stars or in very small groups. We compare this with several other
regions of massive star formation, and with theoretical predictions.

\subsection{Comparison with other regions of massive star formation}

(1) The Orion Nebula Cluster contains about 1600 optical stars with a
total mass in excess of $10^3~\Msun$. We have shown in \S~
\ref{sec:test3} that the energy distribution of the ONC is much redder
than that of the $UV$-bright sources in the Bulge of M51. Only if there
was no extinction at all in the ONC, would its energy distribution be
similar to that of the point sources in M51 with temperatures of about
$25~000$ K.  However, the hotter M51 sources have a steeper energy 
distribution (equivalent to a higher temperature) than the ONC without 
extinction. This shows that these M51 sources have fewer (or no)
low mass stars than the ONC.

 (2) Brandl et al. (1995, 2001) have shown that mass segregation 
has occurred
in the core of the young compact LMC cluster R136a during the 
first few Myrs. The massive stars
are more strongly concentrated towards the center than lower mass
stars. Stars of $M > 25 \Msun$ are concentrated within a core 
radius of 0.1 pc, which is five times smaller than the core radius
of the whole cluster. Could our blue point sources be the cores of
clusters? If the cluster R136a was in M51, we would not
observe the mass segregation and the photometry of the point source
would include (almost) all stars in that cluster. So the ``hot''
energy distribution of many of the sources is not due to mass
segregation, unless the clusters in the bulge of M51 disperse on a
very short timescale of a few Myr.

(3) The bulge point sources in M51 can also be compared with the
clusters that formed due to the interaction of the Antennae galaxies
(NGC 4038/4039). Whitmore \& Schweizer (1995) and  Whitmore et al.
(1999) identified point sources with magnitudes in  the range of $-14 <
M_V < -6$. The  brighter ones are young globular
clusters, and the lower luminosity objects could be stars or small
clusters. This indicates cluster 
formation over the full mass range from $\sim 10^2$ to $10^6$
\Msun. This is different from the situation in the bulge of M51 where
mainly isolated stars are formed. 

\subsection{Comparison with star formation in bulges of other spiral galaxies}

 (1) The inner part of our Galaxy, within a few pc from the nucleus, 
has several very young clusters  with
ages less than 4 Myrs, that contain massive stars.  The best studied of
these are the  so-called ``Arches'' and ``Quintuplet'' clusters at a
distance of  about 30 pc from the galactic center. 
However, these clusters have masses of $1 \times 10^5$ and $1 \times
10^4$ \Msun\ respectively (Figer et al. 1998, 1999) which is much 
more massive than most of the  bulge point sources in M51. 
Rich (1999) has suggested that the star formation near the Galactic
Center may favor the formation of massive stars because tidal 
forces may disrupt
the clusters quickly, perhaps before the low mass stars are formed.

 (2) A detailed study of the stellar population in the inner 
Galactic bulge, in Baade's
Window, at a projected distance of about 400 pc from the center, shows that 
there is no evidence for the presence of luminous young stars, 
with an age shorter than a few Gyrs (Frogel et al. 1999; Frogel 1999). 
This is different from M51 where we do find the young bulge sources
at Galactocentric distances of several hundred parsecs.

 (3) Ichikawa et al. (1998) have shown that the mass-to-light ratios
of bulges of 9 spiral galaxies varies between 0.12 and 3.0 in the
J-band. This corresponds to stellar populations of ages 
in excess of 1 Gyr.

 We conclude that there is little or no evidence for very young stars
in bulges of Galaxies other than M51, apart from the very center
(within 20 pc) of our own Galaxy.

\subsection {Predicted massive star formation in the bulge of M51}

What are the conditions for the formation of massive stars outside
clusters? Norman \& Spaans (1997) and Mihos, Spaans \& McCaugh (1999)
have studied the conditions for the formation of massive stars. They
have suggested that isolated massive stars can be formed in clouds in which
H$_2$, [OI] 63 $\mu$m and [CII] 158 $\mu$m are the dominant coolants. 
This occurs  for instance in
the vicinity of a hot ionizing source in a region with an optical depth
$A_V \le 1$ so that CO is dissociated but  H$_2$ is protected by
self-shielding.

It is very well possible that these conditions are met in the  bulge of
M51: \\
(a) The nucleus of M51 contains a starburst cluster  with an initial 
mass of about $2 \times 10^7$ \Msun\ within the central 17 pc (Paper
I). The UV-radiation field of such a cluster,  a few $10^8$ years after
the starburst  can be calculated with cluster evolution models
(Leitherer \& Heckman 1995).   This yields a UV radiation field
strength at a distance of about 200 pc of $10^2$ to $10^3$ in units of
the mean radiation field in the Galaxy. Models calculated 
by one of us (MS) of the thermal and
chemical structure of interstellar clouds that have 1 or 2 magnitudes
of visual extinction under these conditions, show that CO is largely
destroyed and that  the [OI] fine structure line at 63 $\mu$m is the
dominant coolant (Spaans, Private Communication).

(b) The destruction of CO in clouds of small extinction 
and the increased photo-electric heating will
result in relatively high cloud temperatures of 300 to 2000 K. This
will lead to high a Jeans mass for gravitational contraction  and the
McKee criterium of $A_V \sim 6$ (McKee, 1989) to sustain star formation
is not likely to be satisfied for the bulk of the gas. Since the low
extinction causes any stellar source to induce disfavorable conditions
for further star formation in its vicinity, isolated patches of forming
stars are the natural state.

This suggests that the formation of isolated 
massive stars in the Bulge of M51 could be 
due to the luminous central source, the low dust content,  and the
resulting small extinction. (The star formation under the conditions of
the Bulge of M51 will be described in more detail in a forthcoming
paper by Spaans et al.)


\section{Conclusions}

We have studied bright point sources in the bulge of M51 with the
HST-WFPC2 camera in 6 filters in the wavelength region of  2500 to 8200
\AA. The results can be summarized.

\begin{enumerate}

\item We found 30 point sources in the bulge of M51 with $21.38 < V <
26.20$. These point sources appear to occur in strings that follow the
general pattern of the elliptical or spiral-like dust lanes in the
bulge of M51, but they do not necessarily coincide with the dust lanes.

\item The extinction of the point sources, derived by fitting their
energy distribution with those of single star models or cluster
models, is in the range of $0 < E(B-V) < 1.3$. Half of the objects have
$E(B-V)<0.25$. The absolute visual magnitudes range from about $M_V
\simeq-6$ to $-9$.

\item The energy distributions and the distribution of the  point
sources in the HR diagram suggest that most of them, except one or
two,  are stars rather than clusters because:\\
(a) The observed energy distributions better fit 
those of stellar models than those of cluster models.\\
(b) Many of the sources have an energy distribution and a luminosity 
that is characteristic for a single hot  
massive star of $25~000 \le \Teff \le 50~000$ K. \\ 
(b) The distribution of the objects in the HR diagram  follows 
the Humphreys-Davidson luminosity upper limit for stars.
There is no reason why clusters would follow this limit. \\
(c) There is a gap in the distribution of the sources in the  HR
diagram at intermediate temperatures  between $\Teff \simeq 20~000$ 
and 10~000 K.  This is easily explained if the point sources are stars,
because that temperature range is crossed rapidly by the evolution
tracks of  stars. There is no obvious reason why clusters would avoid
this temperature or colour range.

\item The distribution of stars in the HR diagram shows two groups:
(a) the most massive group of initial mass $M_i > 40~ \Msun$ is mainly
blue and hot, because most of these stars will not evolve into  red
supergiants during their evolution.
(b) the group with $12 < M_i < 25~ \Msun$ is mainly red and cool because
stars in this mass range are below the detection limit during 
their main sequence phase.
 
\item The current star formation rate in the bulge of M51 in the mass range of
$12 < M < 200$ \Msun\ is $\sim 5 \times 10^{-4}$ \Msunyr.  Correcting for
the possible presence of lower mass stars, down to 1 \Msun,  
increases the star
formation rate by a factor 3.4 if we adopt an initial mass function
of slope $\Gamma = -1.35$ (Salpeter's value) and by a factor 1.5 if we
adopt $\Gamma=-0.65$ (the value for the clusters near the Galactic center).

\item 
The total amount of neutral H in the bulge is about $4
\times 10^5$ \Msun\ and the gas-to-dust mass ratio 
about 150. The
 current star formation rate of about  $2 \times 10^{-3}$ \Msunyr\
can be sustained for about 2 to $4 \times 10^8$ years before all the
gas is consumed. This suggests that this form of massive star formation
in the bulge is fed/triggered by the interaction with the companion
galaxy, whose closest approach was estimated to be about $4 \times 10^8$
years ago.

\item These results show that under the conditions that exist in the
bulge of M51, separate massive stars can form outside clusters
or in very small groups. This
agrees with the predictions of Norman and Spaan (1997) who argued that
the formation of massive stars is favoured in regions  near a hot
source (the core of M51) and small optical depth of $A_V \le 1$ so that
CO is dissociated but H$_2$ survives due to self-shielding. This may
resemble the star formation in the early Universe, when the CO content
and the dust content were low due to the low  metallicity.

\end{enumerate}


\section{Acknowledgement}
H.J.G.L.M.L. is grateful to the Space Telescope Science Institute for
hospitality and financial support during various stays. We like to
thank Ron Allen, Don Figer, Massimo Robberto and  Brad Whitmore  for
useful and stimulating discussions about 21 cm observation of M51, the
clusters near the Galactic center, the Orion cluster and star formation
in the Antennae galaxies. Support for the SINS program GO-9114 was
provided by NASA through a grant from the Space Telescope Science 
Institute which is operated by the Association of Universities for
Research in Astronomy, Inc., under NASA contract NAS5-26555.

\end{document}